# LOCATIONS AND STRENGTHS OF SECONDARY RESONANCES LYING WITHIN FOUR FIRST-ORDER MEAN MOTION RESONANCES

Fred A Franklin and Paul R Soper
Harvard-Smithsonian Center for Astrophysics

## I. Introduction

This paper presents a numerical study of secondary resonances that are imbedded within four mean motion resonances, hereafter labeled mmr. Secondary resonance, the concept and importance of which were carefully discussed by Lemaitre and Henrard (1990), refers to a commensurability between apsidal and libration frequencies. Our plan here is to locate their positions and determine their strengths inside the three first-order mmrs in the asteroid belt: 2/1, 3/2 and 4/3, where the integers refer to the period ratios, Jupiter/asteroid. Later we shall also consider the 1/2 mmr, KBO/Neptune, in the Kuiper belt. We emphasize the word 'positions' because chaos, arising when the apsidal and libration frequencies are equal also occurs, though in milder amounts, in a number of 'overtones' [e.g. ones like 3:1, 8:3 and even higher ratios as well]. [Note that we use a slash when referring to mmr ratios and a colon for those of secondary resonances.]

Inasmuch as chaos and orbital instability are generated by an overlap of resonances, in this case by an mmr and a secondary resonance, a study of the extent of secondaries aids in understanding why low eccentricity orbits lying in the 2/1 mmr for example are absent—why lifetimes of hypothetical bodies placed there are typically $O(10^8)$ yrs (Franklin, 1996). Section III introduces the topic of the less extreme depletion that obtains at the 'exterior' 1/2 mmr with Neptune where models show that nearly one-half of bodies captured at 1/2 during Neptune's earlier migration are ejected over a time of $4.6 \times 10^9$ yrs (Franklin and Soper, 2012). A comparison of the origin of chaos induced in these two somewhat similar mmrs has therefore a certain interest. We begin the following section with a study of the structure of the 4/3 Jovian mmr because, unlike 2/1 and 3/2, where some information already exists (Lecar et al., 2001), it provides a fresh example and one located close to Jupiter where only a single asteroid, (279) Thule, has been discovered. It has also proved a useful assist to find that the secondary resonances located there are generally well separated and therefore quite identifiable, not at all the case at 2/1.

## II. Secondary resonances in the asteroid belt

To map secondary resonances in the asteroid belt we use a planar model, integrating with a symplectic routine (Holman and Wisdom, 1993) and kindly made available to us by Matt Holman, adapted to include just Jupiter [J] and Saturn [S].

The two planets start from initial conditions of 1950.0 and the subsequent development of their a, e and ω [apsidal] variations very closely follow their 3D behavior. All massless test bodies start at pericenter, in conjunction with Jupiter. This does lead to a slightly special case that we'll comment upon later. We measure chaos by computing a Lyapunov time, T(L), (cf. Murray and Holman, 1997) determined by following the separation of two nearly identical orbits. 'Nearly identical' means objects having the same orbital elements, but a $10^{-6}$ deg. difference in initial orbital longitude. The growth of this difference provides the abscissa in plots like those of Fig. 1. We do not attempt to calculate precisely very mild chaos and consider orbits with T(L) greater than $10^5$ P(J) to be set aside for our purposes as 'regular enough', but those with log T(L) < 4 as markedly chaotic.

Figures 1-4 map the chaotic structure, owing to secondary resonances, inside the 4/3 mmr at a semimajor axis, a(o), chosen to mimic the orbit of (279) Thule in a 2D representation. Varying the initial eccentricity, e(o), the ordinate on these figures, provides a means of scanning 4/3 and the result produces a fit that matches both Thule's observed proper eccentricity, e(p) = 0.0705 and its apsidal/libration period ratio, P(ω) /P*(l) = 39.003/14.255 [in P(J)] = 2.736, arguing in Fig. 2 that Thule's orbit is very regular in the 2D model. Inasmuch as Thule's inclination remains < 2.65 deg., this regularity seems in little doubt. The initial a(o), chosen as 0.830 [a(J) = 1.0] leads to a semimajor axis range during libration of 0.0215 or 0.112 AU that is the same as that of the real Thule. This range corresponds to a libration amplitude near 30 deg.

We do need to clarify the apsidal/libration period ratios entered on the figures: the (*) attached to P*(l) arises because it is one half the period of libration as it is often defined. This latter quantity measures the elapsed time for a complete oscillation of conjunctions of Thule with Jupiter, starting, say, from the former's pericenter, to one elongation, back through pericenter to the other elongation and returning to pericenter. During this time, the semimajor axis, a, [and eccentricity], goes through two complete cycles, having a maximum at pericenter and minima at each elongation. Since we have used the well-defined oscillations in a to measure libration frequency, the appropriate choice seems to be to plot P*(l) = ½ P(libration). A further defense leans on the fact that semimajor axis [and its variation] is a more basic quantity and perhaps also because P(ω)/P*(l) approaches 1 as e(o) --> 0, cf. Fig. 5. [Obtaining P(lib.) itself is doable enough, but more of a nuisance as it also requires following Jupiter's motion in fine detail.] Our choice means that the 2:1 ratio at e(o) = 0.1275 in Figs. 1 and 2 would be labeled 1:1 had we used P(ω)/P(l), 8:3 would become 4:3 etc and the likely profusion of many ratios lying near P(ω)/P(l) = 1 [e.g. ...3:4, 4:5 ... 1...5:4, 4:3..] are now less obvious in plots using P(ω)/P*(l). But more to the point, using our adopted scheme means that we can widely represent the secondaries shown in Figs. 2-4 by 2n/3 and (2n + 1)/3 for n < 8.

Figure 1 introduces the 4/3 mmr, providing an overall summary, and shows among numerous entries a wide range of (quite) regular and clearly chaotic orbits. The former, those showing log T(L) > 5, are plotted at log T(L) = 5.2, while the four circled cases have escaped—in times < 15,000 P(J) for the two with e(o) near 0.05 and at 78,000 P(J) for the one at e(o)= 0.32.  Orbits of 6 bodies with 0.03 < e(o) < 0.05 [not added to Fig. 1] all escaped in times < 11,000 P(J).  Another escapee, initially with e(o) = 0.0955 and log T(L) = 2.42, departed at $2.8 \times 10^8$ yrs, but two others, marked by small arrows on Fig. 1, at e(o) = 0.073 and 0.129 with log T(L) = 2.66 and 2.63 remained in an apparently stable libration when an integration was stopped at $10^9$ yrs.

The next three figures present greater detail: Fig. 2 shows the low- to-moderate e(o) region, 0.11 to 0.17, that includes a number of secondary resonances, particularly 2:1 and it also locates Thule.  Finding ones like 11:3 was something of a surprise—evidently they are not confined to small number ratios, though the generated chaos is not severe and of limited extent.  Here and on other figures, the > < symbols indicate the e(o) range over which orbits with the indicated P($\omega$)/P*(l) ratio still can occur.  This range is essentially due to variations in P($\omega$) that act to widen a resonance and very little from those in P*(l).

Figure 3 provides detail over the range 0.143 < e(o) < 0.153 that includes the 3:1 and 10:3 secondary resonances.  The behavior at 3:1 seems curious: mildly chaotic orbits on either side of a quite regular one at/near the exact ratio, a condition that also appears at 4:1 and elsewhere.  We suspect, but as yet cannot prove, that this is the consequence of an initial condition that starts all integrations with the asteroid exactly at pericenter when in conjunction with Jupiter.  This procedure could lead to a periodic solution that retains a certain coherence at least for a time.  Such behavior seems related to what happens at e(o) = 0.1585 and shown in Fig. 6: approximate regularity up to a time  of $1.4 \times 10^5$ P(J) and a rapid onset of chaos afterwards.  It is this rise, usually occurring much earlier, that we use to obtain log T(L).  The just mentioned limitation does not compromise estimates of the likely width and depth of a resonance.

In Fig. 4 we complete a survey of 4/3 by moving to higher e(o).  The fact that the region ~0.22 < e(o) < 0.26 shows log T(L) remaining near 3.6 leads to the following speculation: at e(o) smaller than ~0.20, resonances develop some chaos but only over a narrow discrete ranges that remain independent while to the right of 0.26, they are too weak—of too high order—to have more than a small effect. Within the interval itself, 0.22 < e(o) < 0.26, the weakening resonances are wide enough to overlap, as the > < symbols suggest, and the chaos stays quite continuous and constant.  The ratios listed for e(o) > 0.2 are meant only as positional guides, not as the only identification so we could equally well have used others.  As e(o) reaches ~ 0.3, close Jovian approaches occur, leading to increasing instability and finally an escape at e(o) = 0.32, at T = 78,000 P(J). Until we can address secondary resonance on a theoretical basis, such remarks are only tentative ones.  For a firmer conclusion, we can state that the 4/3 mmr includes a region of e(p) from ~0.06 to ~0.13, extending with some omissions at and near 2:1 even down to e(p) = 0.03 where a great many quite regular orbits exist and even many of the chaotic ones are only mildly so.  Or, in other words, Thule is safe, as a $2 \times 10^9$ yr integration of 10 bodies at/near Thule (Lewis, private communication) supports and this is the place, including the regions of mild chaos, where we would expect, on stability grounds alone, to find others like it. The failure to discover additional ones probably owes its explanation to the difficulty of capture in a region so close to Jupiter (cf. Franklin et al., 2004)

Our next step compares the survey of the 4/3 mmr plotted in Fig. 1 to ones for 3/2 and 2/1. Figure 7 for 3/2 is not all that different: a number of identifiable secondaries with an extended region between ~0.048 < e(p) < 0.180 with many very regular orbits. Here at 3/2, most can be represented by n/2, with n < 14, suggesting there is a characteristic difference between this mmr and 4/3 where the secondaries are measured with a denominator of 3. In comparing Fig. 7 with the observed distribution of the Hilda minor planets [the given name of bodies librating at 3/2], we find that about 40% of the 185 with well-measured orbits lie in this region, with the remainder, having 0.18 < e(p) < 0.28, moving in somewhat chaotic orbits. Among the 185, the largest e(p) is 0.281. Fig. 7 points to an apparent anomaly: at e(p) smaller than ~ 0.1 only 3 asteroids with accurate orbits and only 3 or 4 additional suspects with e(p) < 0.1 are known. But Fig. 7 argues that there is no reason based on a dynamical instability to account for the observed population shortage for 0.048 < e(p) < 0.1. This short-coming prompted Franklin et al.(2004) to propose an evolutionary mechanism that develops readily from Jupiter's early inward migration.

As in the example at 4/3, we again speculate that log T(L)'s again of ~ 3.6 for e(p) > 0.18 arise from an overlap of separate secondary resonances. A secular resonance located here is an unlikely source of chaos: even at e(o) = 0.350, or e(p) = 0.275, $P(\omega)$ is 680 P(J) which is too short to be involved with any secular resonance nor can a large libration amplitude contribute as they remain < 50 deg. for 0.2 < e(p) < 0.3. [The amplitudes increase with declining e(o).] However, Morbidelli and Moons (1993) show that secular resonance can become important as e(p) rises much above 0.28 and it is the likely reason why the three orbits at e(o) = 0.430, 0.435 and 0.440 escaped—at times of 350,000, 67,000 and 23,000 P(J). Two other escapes at low e occurred at e(o)'s of 0.008 and 0.024 at $1.6 \times 10^5$ and $1.9 \times 10^7$ P(J).

The most interesting case, that of the chaos in the 2/1 mmr, summarized in Fig. 8, is drawn to the same scale as Fig. 7 and nearly the same as for 4/3 in Fig. 1. The contrast with the other two is striking: no likely regular orbits until e(p) rises well above 0.5 and severely chaotic ones for all e(p) < 0.2. Since we've considered only one value of a(o), these numbers will vary somewhat, but to no important extent as we have verified in several examples. The escape of bodies of low eccentricity from 2/1 has already been clearly noted in times of $O(10^8)$ yrs. Now, for the 2D integrations, seven bodies, the circled ones, escaped in times ranging from $1.2 \times 10^7$ to $1.7 \times 10^8$ yrs. Somewhat curiously, the object at e(o) = 0.085 with log T(L) < 1.94 escaped last. This is not really so surprising because, although there is a clear trend linking short T(L) with more rapid escape, the relation is by no means a precise one [Lecar et al., 2001].

The square brackets in Fig. 8 indicate that, owing to large variations in $P(\omega)$ and much lesser ones in P*(l), secondary resonances, in addition to the one marked at the base of each dotted line, will temporarily occur everywhere to the left of 7:1. The listed ones represent extremes so that the range may also include resonances lying between them. Increases and decreases in P($\omega$) allow this condition to repeat: the 2:1 resonance for example appears first at the location marked 3:1, an average value, then later near 10,000 P(J) and again at 14,000. Figure 9 clearly shows the resulting chaos that the presence of 2:1 will generate. In fact 2:1 does occasionally make its appearance at least up to e(o) = 0.09.

To the right of 7:1, the individual secondaries begin to overlap [though there are none of the excursions shown at lower e(o)] because of the narrowing separation between them as e(o) and P($\omega$) increase [cf. Fig. 5] -- much as we noted in considering the 4/3 mmr. We see this overlap, which here includes only resonances designated by whole number ratios, as a 'weaker relative', but still similar to the 'travelling' resonance behavior seen in the low e(o) area. For a quick circumstantial summary, we can say that the greater chaos at 2/1 is fueled by the much larger variations, especially in apsidal motion, prevailing there.

### III. A Case in the Kuiper Belt

This section moves attention to the 1/2 mmr with Neptune at 47.76 AU at the border of the Kuiper belt. Its dynamics provide an intriguing commentary on the history of that planet's outward migration. Franklin and Soper [2012] found that, with only a single exception, all higher order mmr outwards of 1/2 captured and retained a few bodies as Neptune's orbit expanded, but the more efficient capture at 1/2 was accompanied by a later escape from this resonance—to such a degree that nearly half of those captured, 31 of 69, were ejected during 4.6 x $10^9$ yrs. [The exceptional case was at 5/11 where 2 of 6 captured bodies both escaped after about 3 x $10^9$ yrs.] The question then arises: is chaos at 1/2 generated by secondary resonances lying with it severe enough to account for this instability? A quick look when we estimate or determine P($\omega$) and P*(l) might at first seem discouraging. P*(l) scales inversely as the 2/3rds power of the mass of the [planetary] perturber so that the ~ 35 P(J) period at 2/1 with Jupiter presumably would rise by something like a factor of 7 at 1/2 with Neptune. In fact, our calculations find the range 250 < P*(l) < 450 P(N). But values of P($\omega$) are much longer: 15,000 to 45,000 P(N) for 0.03 < e < 0.15 at 1/2.

Despite the large ratio of these periods, Fig. 11 that plots $\omega$ vs. time for e = 0.17 provides a measure of hope as it indicates the presence of another much shorter periodic variation. Its period remains very close to 300 P(N) for 0 < e < 0.25, declining slightly to 280 P(N) at e = 0.01. This term does not represent a response to any resonant perturbations as it is always present, not just at a(1/2) = 1.5874 [a(N) = 1.0], but also over the range 1.4 < a < 1.8. The comparable magnitudes of P*(l) and the short period term in P($\omega$), hereafter labeled P*($\omega$), led to the survey at 1/2 in Fig. 10. It generally follows the approach used to produce Figs.1, 7 and 8 with two differences: 1) integrations now include the four major planets and, though still in a planar approximation, extend for at least 40,000 P(N) or 6.6 myr and 2) to maximize chances for a stable libration, the initial conditions are now set to ensure that a test body lies at its apocenter in conjunction with Neptune. We chose to emphasize the differences by changing the notation to $P^*(l)/P^*(\omega)$ rather than the earlier $P(\omega)/P^*(l)$. Figure 10 can now omit noting values for e(p) because the eccentricity forced by Neptune, thanks to its nearly circular orbit, is nearly zero so that, as checked, e(p) lies very close to e(o).

Figure 10 does closely resemble its earlier counterparts, especially by showing the marked chaos for e < 0.1. As before at 2/1 there are repeated episodes when several period ratios at a given e(o) occur. The recurring ones lie at values of $P^*(l)/P^*(\omega)$ between 1:1 and 2:1, with 3:2 a frequent contributor. For times when a commensurability develops, usually for a few thousand P(N), chaos is indicated by an exponential rise in the longitude separation of the two test bodies while at other times any increase is barely detectable, much as was shown by the case in Fig. 9. To present a balanced picture, Fig. 10 plots two sets of symbols: filled squares are values of log T(L) averaged over 40,000 P(N), while crosses measure log T(L) over just the times when a secondary resonance is clearly present. The latter therefore mark the maximum level of chaos.

A caveat applies to parts of this discussion: the periodic variation in semimajor axis that we use to determine the libration period is obviously present when e(o) > 0.06; it is identifiable for somewhat smaller e, but becomes undetectable for e(o) < 0.04. The small e(o) cases, however, even as low as 0.01, do develop higher eccentricities, at which times the semi-major axis variation becomes definite and readily measureable. Still, we cannot claim that a libration and not, say, a circulation is present when the eccentricity is, or remains, < 0.04.

Looming ahead is the question whether this degree of chaos is sufficient to correspond to an escape and here is a query to which we cannot provide a clear answer. Values of log T(L) < 3, which Fig. 10 shows do occur at small eccentricity, do suggest a 'yes' for escape. Although the above limit seemed a likely approximate criterion near Jupiter, as we review in part IV, a somewhat higher value may apply in the weaker dynamical environment in the outer solar system beyond the orbit of Neptune. Finding only very slight signs of chaos at 1/2 would have left the 'observed' escape of bodies as being quite mysterious. A search at 1/3 and 2/5, where our earlier paper found no escapes over $4.6 \times 10^9$ yrs, somewhat strengthens the case for escape at 1/2 because no significant chaos appeared at either of these two mmr at low eccentricity. Instead we want: At 1/2 when e(o) > 0.11, values of log T(L) remain greater than 5 and fall occasionally below 4 only at e(o) ~ 0.2. We also have found no signs of threatening chaos at the 2/3 mmr at a = 39.43 AU. Our claim, or suggestion, that secondaries do produce the chaos at 1/2 seems therefore very likely though not completely proven and even a quantitative link between the measured chaos, however it is generated, and the slow ejection of bodies is not absolutely established.

## IV.  Final Remarks

Our opening concern in this paper centers on the dynamical status of three first-order mean motion resonances, 2/1, 3/2 and 4/3 in the asteroid belt owing to the influence of secondary resonances, i.e., the many commensurabilities between the apsidal [$\omega$] and libration [l] frequencies.  The three mmr do have one common feature: for small proper eccentricity the [numerical value of 'small' varies among the three] most orbits are extremely chaotic with log T(L) < 2.5, escapes do occur and some happen in times very much less the solar system's age.  The less extreme chaos for 3/2 shown in Fig. 7 at e(o) < 0.04 argues that escapes are much less likely there when compared to the parallel situation at 2/1 and indeed a few well observed Hildas [e.g. (334) Chicago and (1256) Normania] with e(p)'s close to 0.03 are known.  At 2/1 the continuous run of chaos means that we have no clear way of determining to what extent the likely profusion of secondary resonances drives this process.  But as the e's increase toward and beyond the 2:1 secondary, their presence becomes clear at 4/3 so that it's likely enough that they exist at 2/1 as well.

At 4/3 and 3/2, chaos induced by the secondaries at e's beyond 2:1 is quite localized and of limited severity.  It poses no real threat to orbital stability or leads to any depopulation of bodies up to $4.6 \times 10^9$ yrs.  On these grounds it is clear that both 3/2 and 4/3 can at present harbor bodies over the wide ranges of eccentricity that Figs. 1 and 7 indicate.  To elaborate a little further: for e(p)'s over a broad range close to and larger than 0.2, these two mmr, especially 3/2, show a quite continuous number of orbits with a non-threatening log T(L) near 3.6.  Near these higher e's, the rapidly rising P($\omega$) forces the now weakening secondaries to approach each other, becoming more dense and consequently overlapping so as to generate the mild chaos.  What makes this all the more possible is a broadening effect thanks to small variations in P($\omega$) at each secondary.  And what really distinguishes 2/1 from the other two is the far greater magnitude of the variation in P($\omega$) -- to such an extent that the secondaries can wander and recur over a large range of eccentricity.   This seems a clear way of representing and comparing behavior, but it is only that, and deeper explanation as to why this happens is needed.

Some of the above remarks bring up the question of a link between T(L) and time of escape.  The limited data here suggest that orbits with log T(L) less than 3 are at least at risk.  This claim is compatible with some valuable results of Holman and Murray (1996).  They calculated T(L) for all known outer belt minor planets, presumably bodies that have remained from very early times in the solar system.  Of 33 orbits considered, just 7 have log T(L) < 3, the smallest being that of (4236) with log T(L) = 2.48.  These results find a positive response in a comparison of the low e(o) region between Figs. 1, 7 vs Fig. 8.  For the latter that applies to 2/1, the log(T)'s extend to lower values and escapes are the more common.  A related stability concern or potential observational conflict is provoked by the rarity of real asteroids for e(p) < 0.1 at the 3/2 mmr.  Figure 7 shows that only regular or mildly chaotic but non-escaping orbits exist for all e(p) > 0.048.  A rather automatic explanation does follow if Jupiter has migrated inward to its present location by about 0.4 AU, a process that would first efficiently capture and later expel asteroidal bodies of low eccentricity from the 5/3 mmr [where there are none today] at times before any capture into 3/2 would have occurred. This analysis forms the major part of a paper (Franklin et al., 2004) that also estimates the low capture probability at 4/3.

One area of numerical study needs more effort. We have mentioned the help provided by the 4/3 mmr to locate secondary resonances, but even there we lack a clear inventory of their presence near and on both side of 2:1. We plan to use models with a reduced mass for the two planets to clarify his matter and to try to extend the attempt also to 2/1 -- to see, for example, at what Jovian mass they remain separate entities and then the value for which they begin to overlap. There's also the interesting question that we have only found a symptom of: the large $P(\omega)$ variations at 2/1 and resulting marked chaos. As a small step, we do know that chaos is particularly manifest in angular variables so that the large changes in $P(\omega)$ do reflect its severity and influence.

This paper added Section III to look for a possible role for secondary resonance in developing chaos in, and escape from, the 1/2 [Kuiper belt] mean motion resonance with Neptune. The presence of a nearly constant short period term in the apsidal motion—one very much less than the time for a complete apsidal rotation—that equals multiples of the libration period does generate considerable, though variable chaos for e < 0.10. We mention 'variable' because there are fluctuations of a factor of nearly 2 in the libration period, meaning that the occurrence of chaos is intermittent. The efforts discussed in Section III lead us to the definite suspicion, but to no clear proof, that the continuing escape of the bodies once captured into 1/2 much earlier [cf. Franklin and Soper, 2012] might be accounted for by the instability generated by secondary resonances imbedded within the 1/2 mean motion resonance.

# Figure Captions

Fig. 1: Chaotic and more regular orbits at the 4/3 mean motion resonance. Circles mark 4 escaping orbits, arrows 2 that remained for one billion yrs.  We include the apsidal/libration period ratios, P(w)/P*(l) and proper eccentricities, e(p).

Fig. 2: More detail at 4/3, indicating chaos at/near the 2:1 secondary resonance and the location of the apparently stable asteroid (279) Thule, denoted by the vertical line with smaller dashes.

Fig. 3: Greater detail near 3:1. As text discusses, the apparent regularity of the orbit exactly at 3:1 is probably a consequence of special initial conditions.  The > < symbols here and elsewhere measure the limit to which a secondary resonance can extend.

Fig. 4: Secondaries, still at 4/3, with higher e(p).  We suppose that the overlap of weak ones [we plot only even alternates to avoid confusion] generates the trough over a range centered on e(p) ~ 0.17.

Fig. 5: Mean apsidal and libration periods at the 4/3 mean motion resonance.

Fig. 6: A quite rare case of delayed chaos--more details in the text. The rise at T ~ 140,000 P(J) is used to obtain the Lyapunov time, T(L).

Fig. 7: A survey of the 3/2 mmr, including 5 escaping orbits at high and low eccentricity.  Note the similarity to Fig. 1 and the numerous regular orbits for ~ 0.048 < e(p) < 0.1 where observations find very few.  Once again, it seems possible that overlap of several secondaries, owing to small variations in P(w), beyond 11:1 is the source of mild chaos for e(p) > 0.18.

Fig. 8: The striking difference of a survey at 2/1: severe chaos at least as high as e(p) ~ 0.2.  Numbers in [ ] indicate the occasional presence of secondary resonances other than the one marked below.

Fig. 9: A rather typical occurrence at 2/1 where the 3:1 secondary is temporarily replaced by 2:1 near t = 10,000 P(J) and again near 14,000, both marking an onset to chaos.

Fig. 10: A survey with period ratios redefined at the 1/2 resonance in the outer Kuiper belt.  Crosses correspond to log T(L) when secondary resonances occur; filled squares to an average value over 40,000 P(Neptune).

Fig. 11: Apsidal motion for e(o) = 0.17 at the 1/2 resonance with Neptune. We suppose that commensurabilities between the short period term of ~ 300 P(N) shown here and the libration period of comparable magnitude produce secondary resonances and lead to the chaotic behavior shown at small eccentricities in Fig. 10.

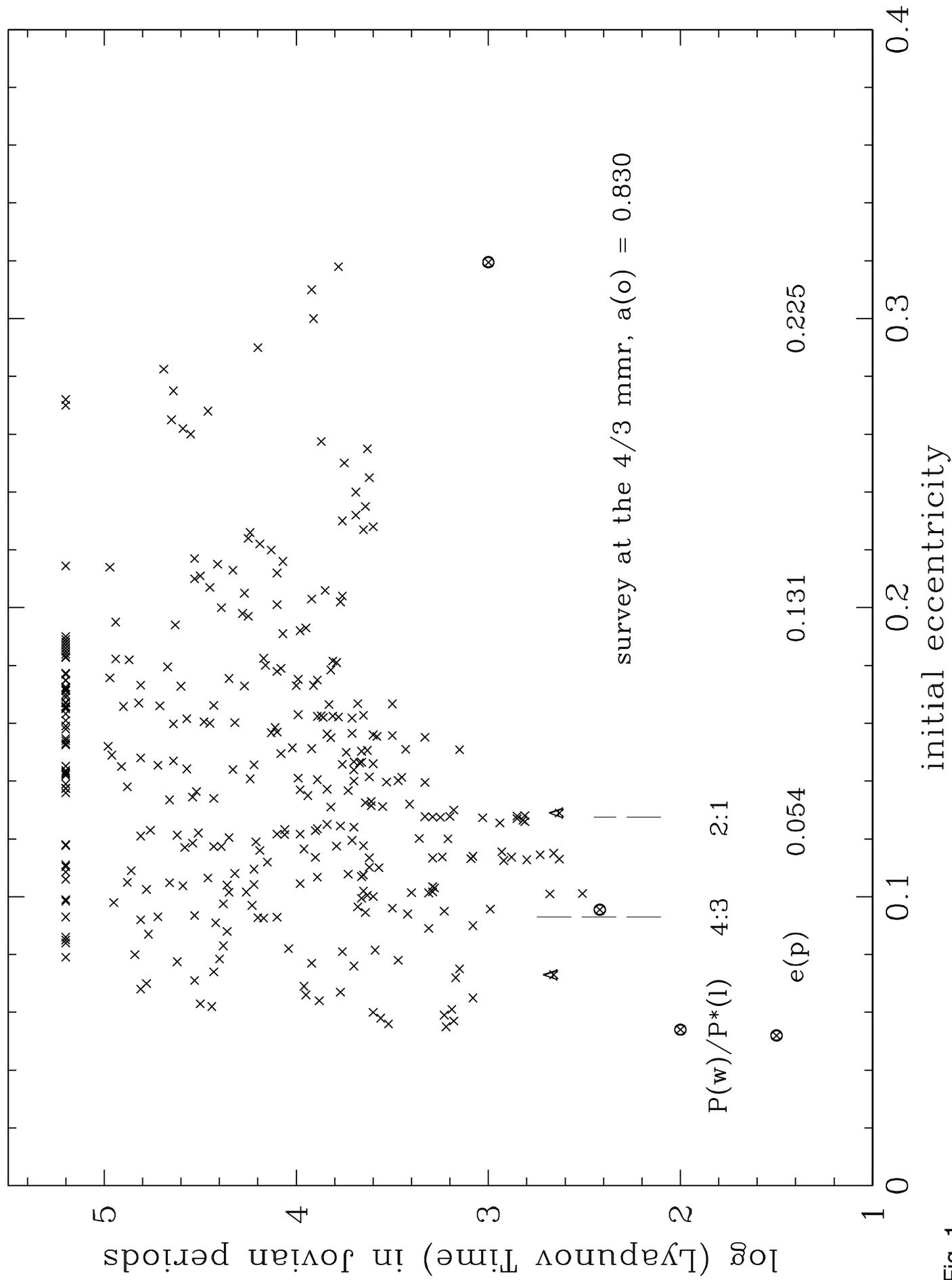

Fig. 1

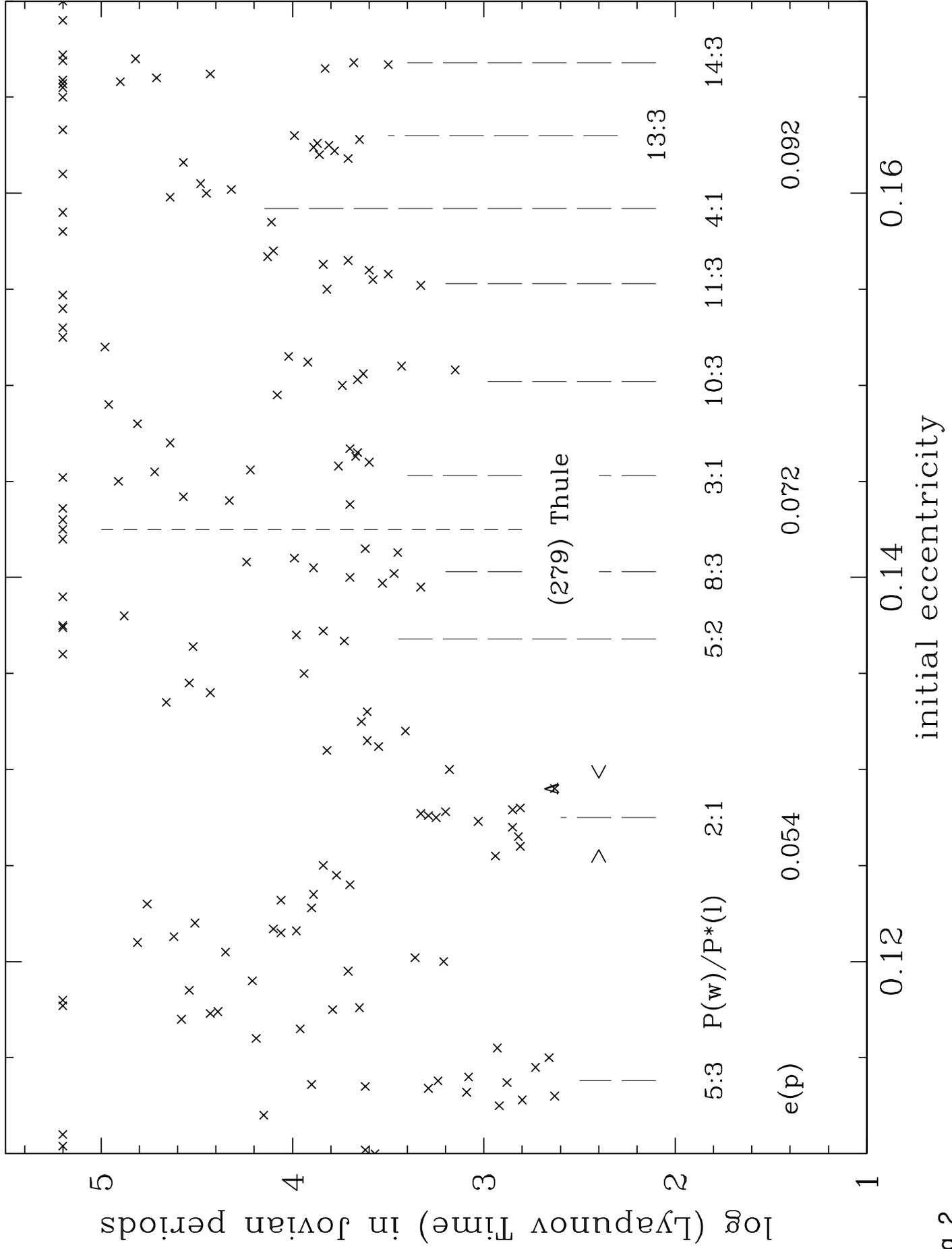

Fig. 2

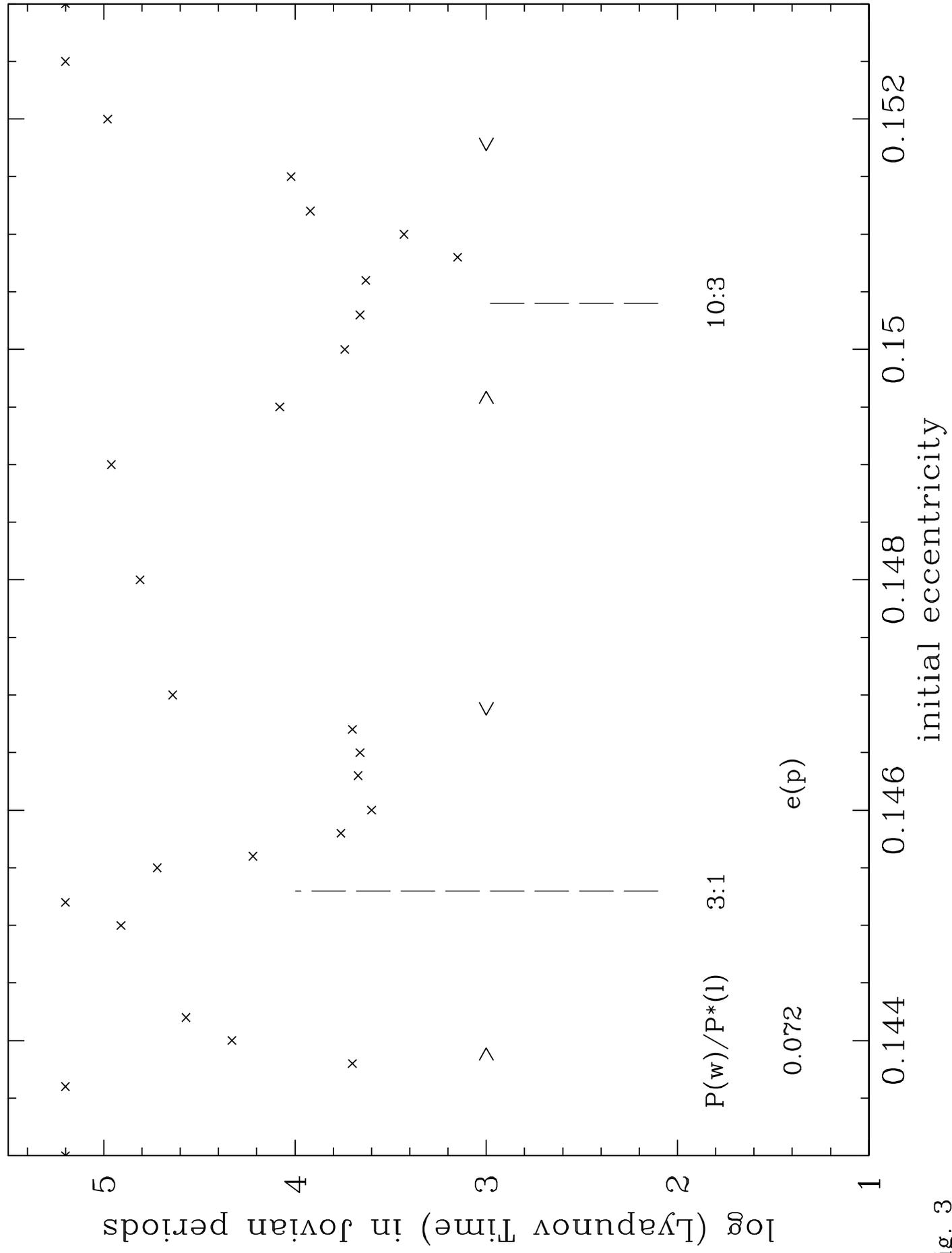
Fig. 3

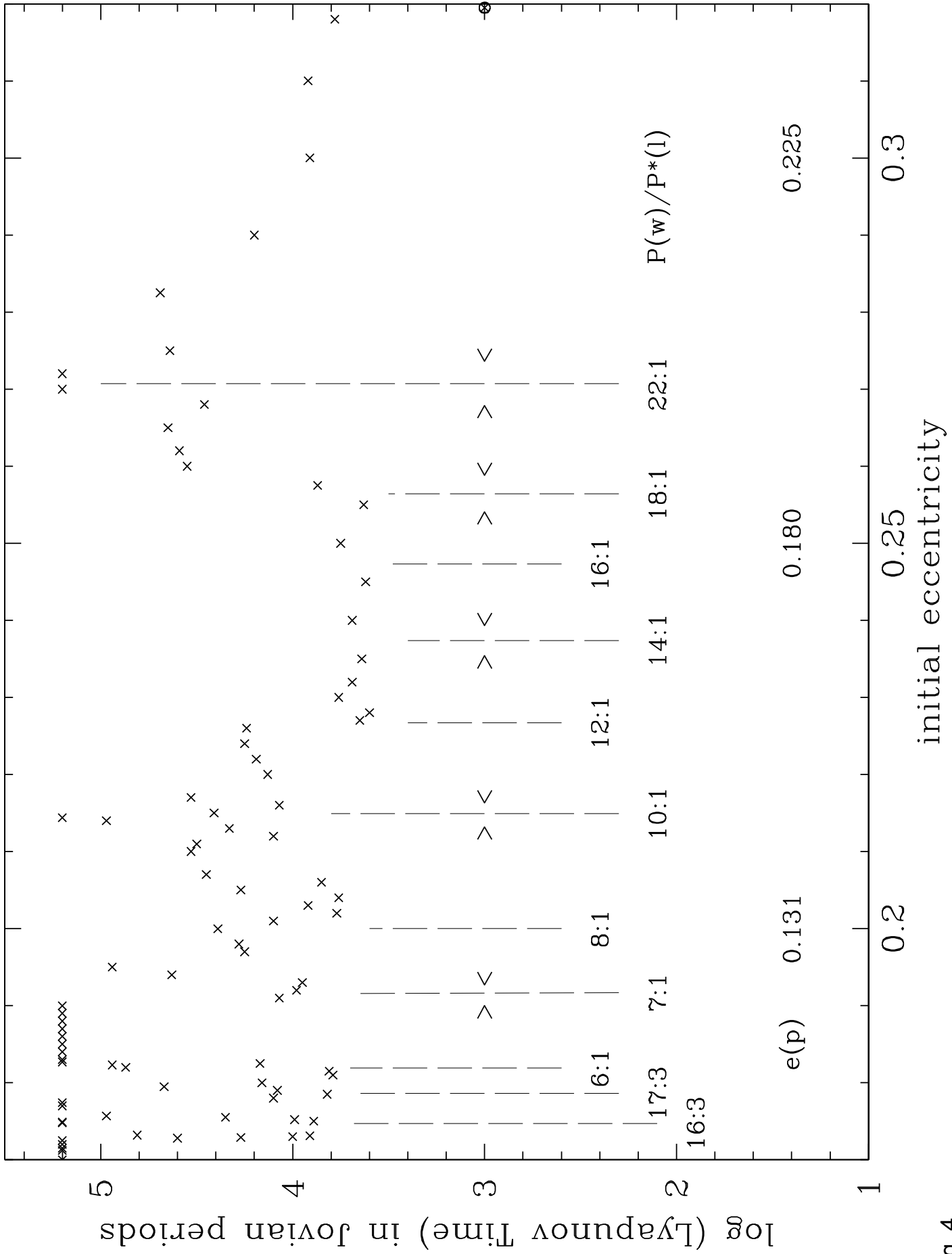
Fig. 4

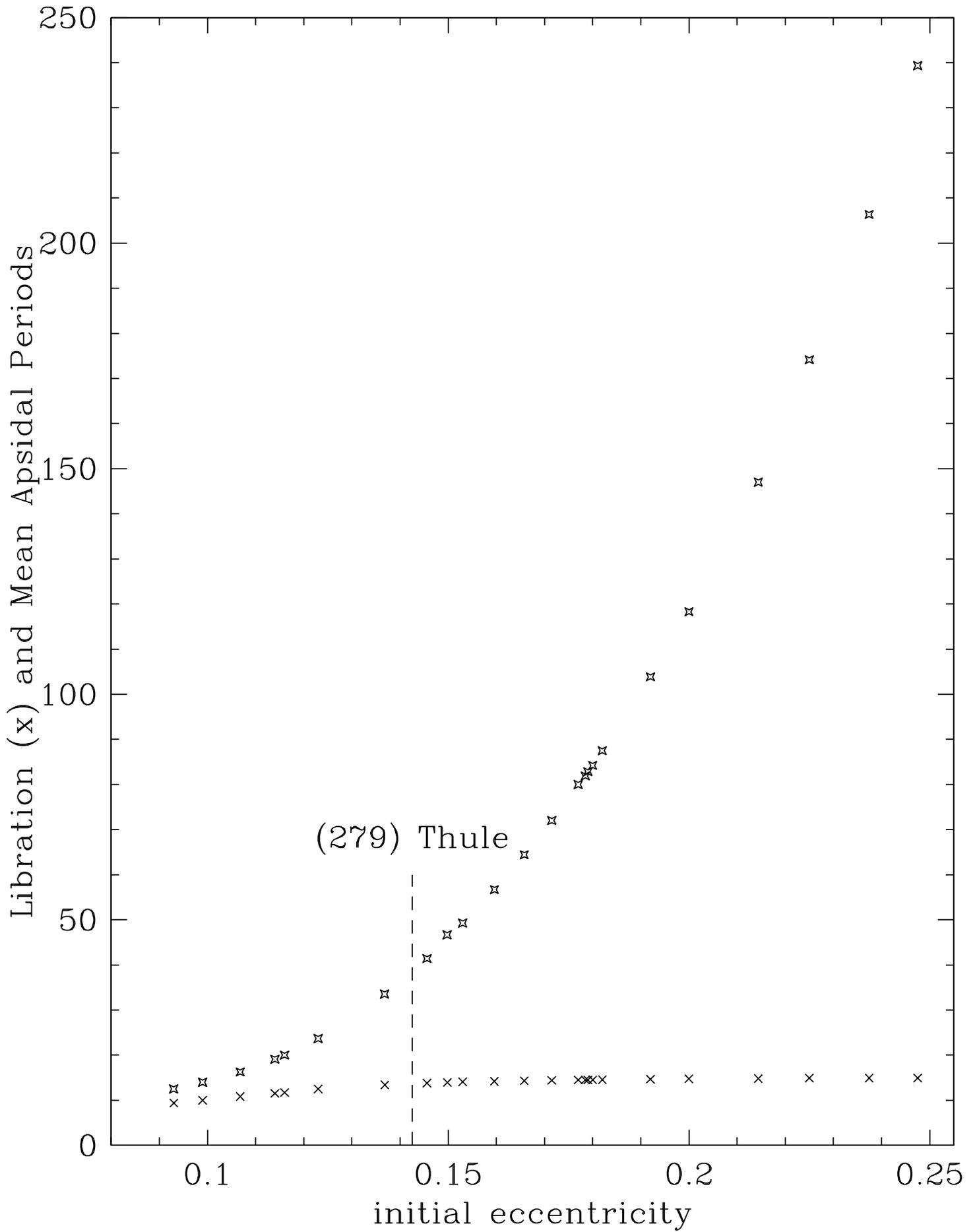

Fig. 5

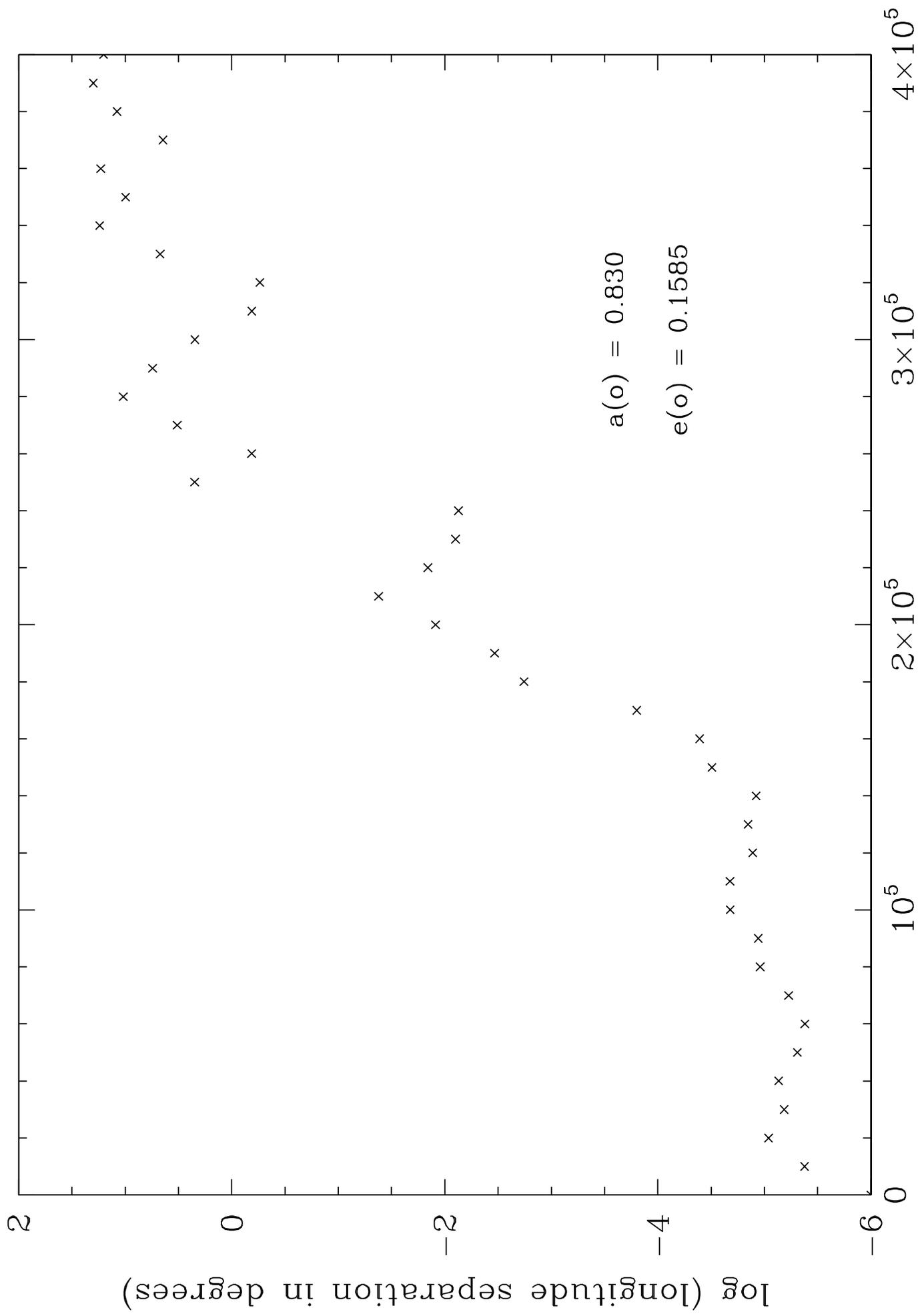

Fig. 6

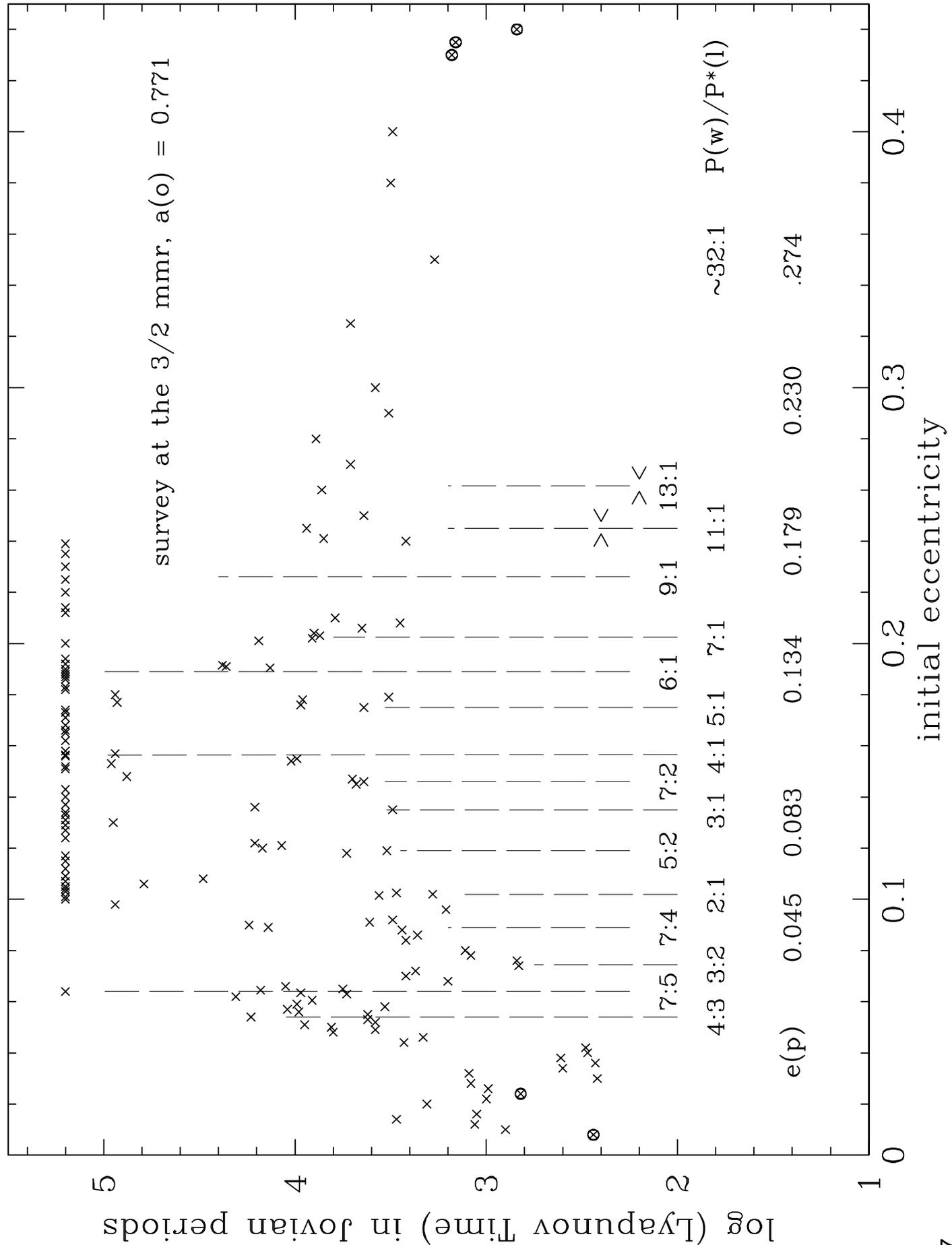

Fig. 7

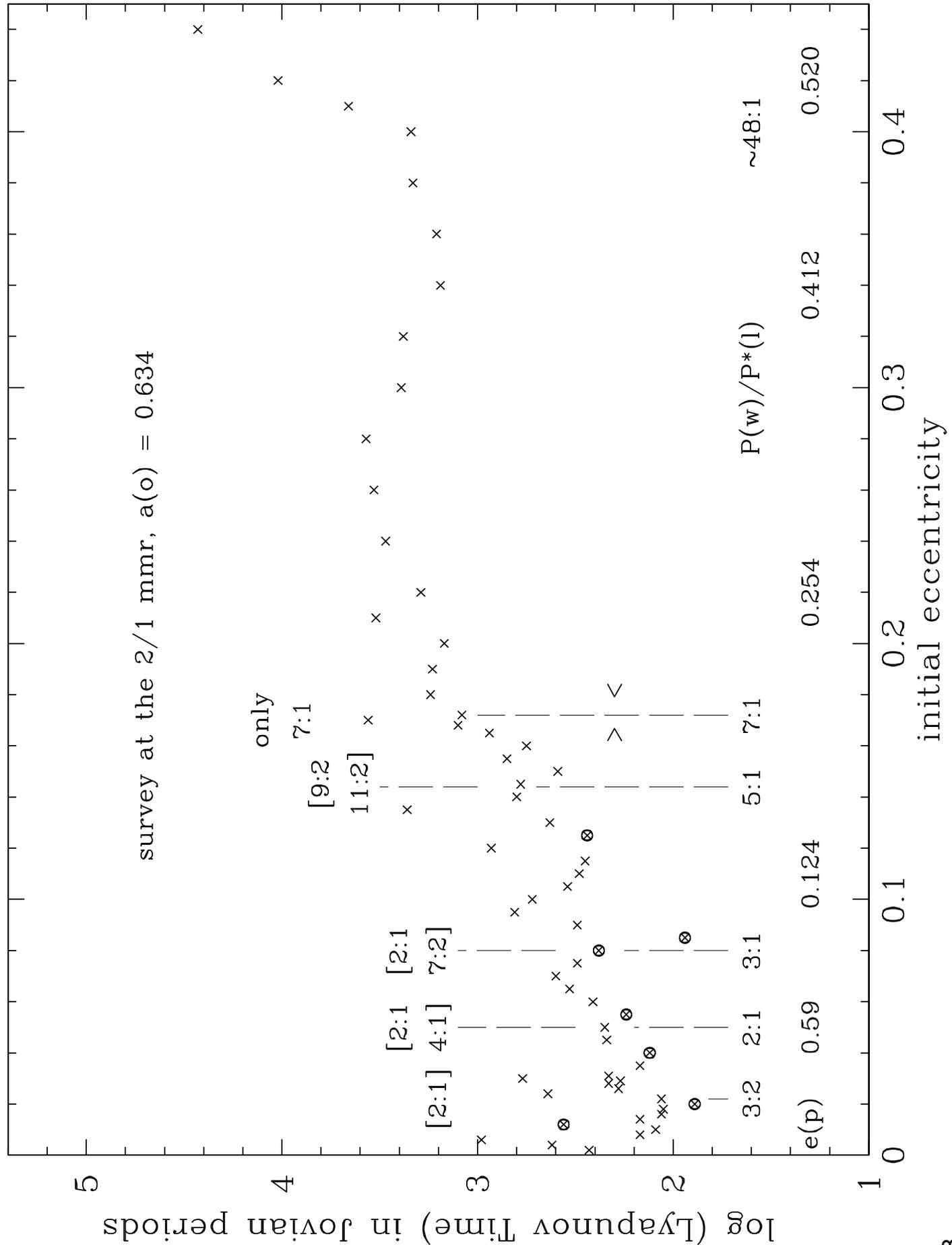

Fig. 8

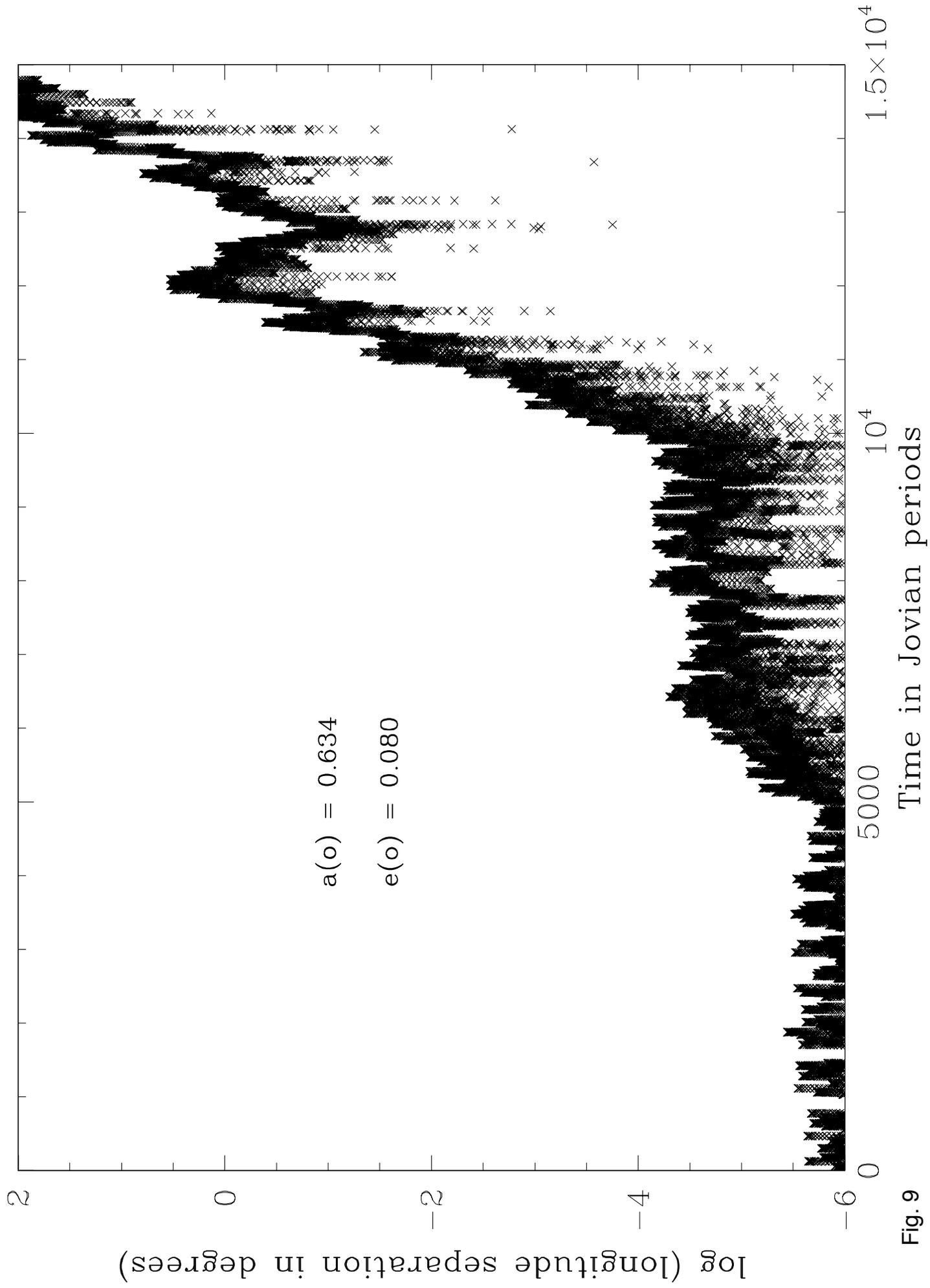
Fig. 9

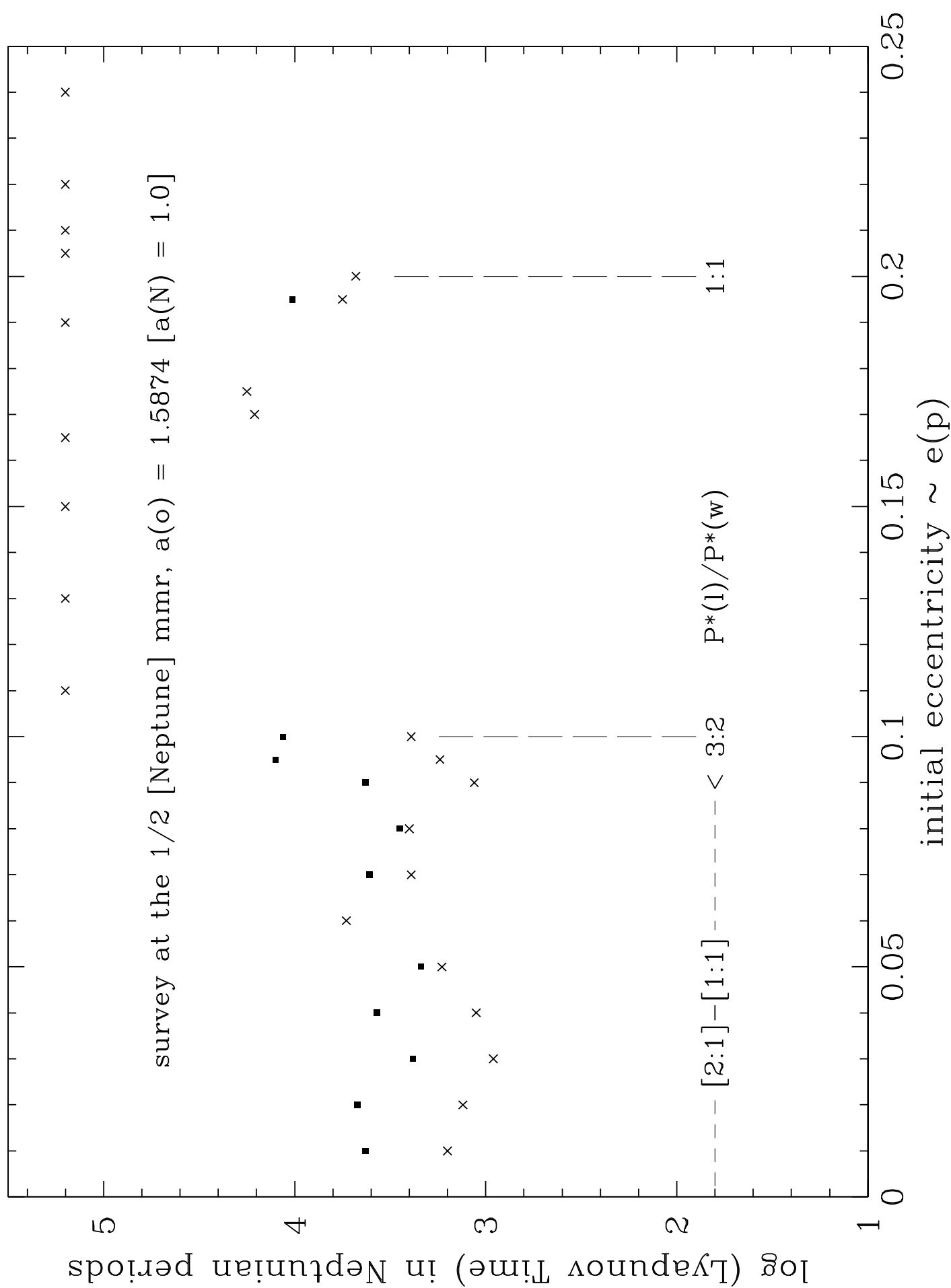

Fig. 10

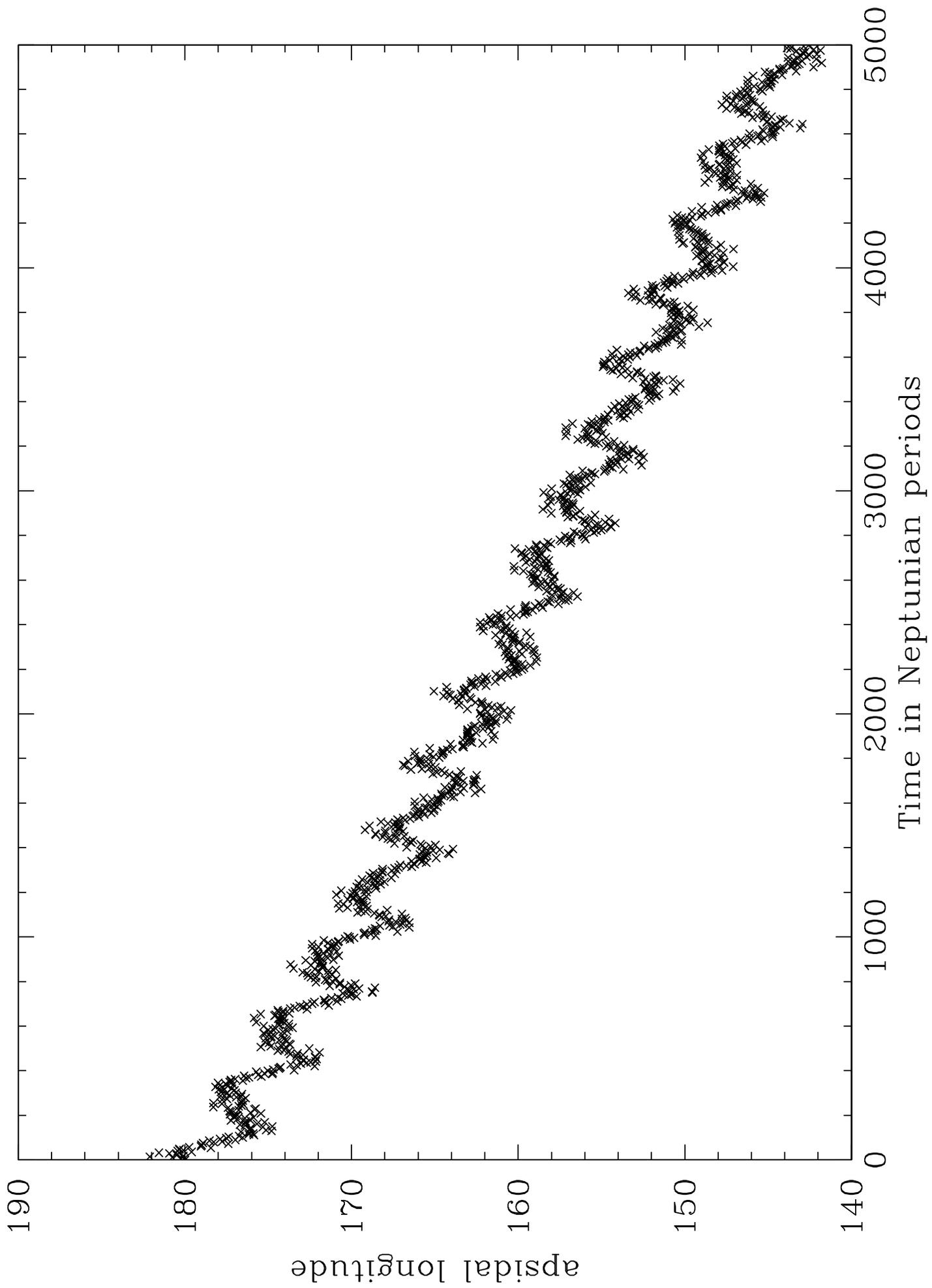

Fig. 11